\documentclass[aps,pra,twocolumn,showpacs,showkeys,superscriptaddress,reprint]{revtex4-1} 
\usepackage{graphics} 
\usepackage{graphicx} 
\usepackage{amsmath} 
\usepackage{color}
\usepackage{soul} 
\usepackage{ulem} 

\begin{document} 

\title{Ground state of a mixture of SU($3$) fermions and scalar bosons} 

\author{J. Silva-Valencia} 
\email{jsilvav@unal.edu.co} 
\affiliation{Departamento de Física, Universidad Nacional de Colombia, Bogotá D.C. 111321, Colombia.}
\affiliation{Department of Mechanical Engineering and Materials Science, University of Pittsburgh, Pittsburgh, PA, USA.}
\author{J. J. Mendoza-Arenas}
\affiliation{Department of Mechanical Engineering and Materials Science, University of Pittsburgh, Pittsburgh, PA, USA.}
\affiliation{Department of Physics and Astronomy, University of Pittsburgh, Pittsburgh, PA, USA.}
\date{\today} 

\begin{abstract}

We explore a system composed of scalar bosons and SU($3$) fermions in one dimension. Considering only local intra- and interspecies interactions, the system is described by the Bose-Fermi-Hubbard Hamiltonian, which is studied using the density matrix renormalization group method. In addition to the common gapless and mixed Mott insulator phases, we identify unknown gapped states, in which bosons couple with some or all flavors of fermions. We unveil different flavor-selective states characterized by itinerant fermions that coexist with an insulator state where the bosons tie with only one- or two-flavor fermions. The states reported here could be implemented in cold-atoms experiments.
\end{abstract} 


\maketitle 

\section{Introduction}\label{sec:intro}

Systems composed of bosons and fermions are nowadays common in many scenarios, offering a wide range of interesting physical phenomena. For instance, the interaction between electrons and phonons leads to Cooper pairing  crucial in conventional superconductivity~\cite{BCS-1957}. Other examples of Bose-Fermi mixtures are the quark-meson models in high-energy physics~\cite{Schaefer-NPA05}, excitons and quasiparticules in semiconductor hetereostructures~\cite{Schwartz-S21,Lagoin-NM23,YZeng-NM23,Gao-NCom24,ZLian-NP24}, and the $^3$He-$^4$He combination~\cite{Andreani-JPCM06}, each one with its own characteristics and relevance.\par 
Currently, ultracold atom setups constitute a common platform to produce and investigate mixtures of bosons and fermions, due to the exceptional control over the physical parameters. This includes interparticle interactions, filling factors, lattice geometry, etc.~\cite{IBloch-RMP08,Esslinger-AR10,IBloch-NP12,Gross-S17}. These capabilities have allowed experimental researchers to achieve several Bose-Fermi mixtures~\cite{Chin-RMP10,Schafer-NatRP20,Baroni-ArX24} and observe intriguing phenomena, such as phase separation~\cite{Lous-PRL18}, effective interactions between bosons mediated by fermions~\cite{DeSalvo-Nat19}, an asymmetry behavior of the speed of sound~\cite{Patel-PRL23,XShen-PRL24}, among others.\par
The theoretical description of a degenerate mixture of bosons and fermions is usually based on the Bose-Fermi-Hubbard model, which considers the kinetic energy of bosons and fermions, as well as local intra-species and inter-species interaction terms. Regardless of whether the internal degrees of freedom are considered or not, interesting predictions have been made, including the emergence of charge density waves, Mott insulators, superfluids, phase separation, Wigner crystals, and spin density waves, among other ground states~\cite{Albus-PRA03,Cazalilla-PRL03,Lewenstein-PRL04,Mathey-PRL04,JSV-JPCM01,Roth-PRA04,Frahm-PRA05,Batchelor-PRA05,Takeuchi-PRA05,Pollet-PRL06,Mathey-PRA07,Sengupta-PRA07,Mering-PRA08,Suzuki-PRA08,Luhmann-PRL08,Rizzi-PRA08,Orth-PRA09,XYin-PRA09,Sinha-PRB09,Orignac-PRA10,Polak-PRA10,Mering-PRA10,Anders-PRL12,Masaki-JPSJ13,Bukov-PRB14,TOzawa-PRA14,Bilitewski-PRB15,Sowinski-RPP19}.\par  
The ground state of a degenerate system of interacting bosons or fermions can be gapless or gapped, the latter being the well-known Mott insulator. In this state, an integer number of carriers (up to one for fermions and larger than one for bosons) can occupy a single site in the lattice, a prediction that has been widely corroborated in several experiments~\cite{Greiner-N02a,Jordens-N08,Schneider-S08,Bakr-S10,Sherson-Nat10,Greif-Science16}. A Mott insulator implies that a commensurability relation is fulfilled between the number of carriers and the lattice, which can be expressed in terms of the density (number of carriers /lattice size) as $\rho^F=1$ or $\rho_B=n$ for fermions or bosons, respectively ($n$ being an integer). Initial studies of polarized bosons and fermions predicted a mixed Mott insulator. Here, the total number of carriers (bosons + fermions) is commensurate with the lattice, satisfying the relation $\rho^F+\rho_B=1$~\cite{Zujev-PRA08}. This mixed Mott insulator was later verified in experiments with ytterbium atoms~\cite{Sugawa-NP11}.\par
Current experiments on Bose-Fermi mixtures involve highly-magnetic atoms, making relevant to consider the internal degrees of freedom of one or both types of carriers. A step forward was taken by considering a one-dimensional mixture of two-flavor fermions and scalar bosons, for which an interplay between phase separation, charge density wave and  superconducting phases of fermions induced by the bosons was reported~\cite{Schonmeier-Kromer-PRB23}. Also, it was shown that the visibility of the Fulde-Ferrell-Larkin-Ovchinnikov state is enhanced as the boson-fermion repulsion grows~\cite{Singh-PRR20}. On the other hand, we reported the existence of the mixed Mott insulator and an unusual insulator where one-flavor fermions satisfy a commensurability relation with the bosons, while the others remain in a gapless phase. This state is referred to as a flavor-selective insulator, and fulfills $\rho_B + \rho_{\uparrow,(\downarrow)}^F=1$~\cite{Avella-PRA19,Avella-PRA20,GuerreroS-PRA21,JSV-RACCFYN22}. Furthermore, a follow-up study unveiled the emergence of other three different insulators, one of them flavor-selective, due to the nearest-neighbor interactions between fermions or bosons~\cite{GomezLozada-ArX23}. Flavor-selective insulators were first identified in the ferromagnetic phase of the Kondo lattice model~\cite{Peters-PRL12,Peters-PRB12} and recently observed in experiments with $^{173}$Yb atoms, where the symmetry of the SU($3$) Fermi-Hubbard model was broken in a controlled way~\cite{Tusi-NP22}.\par 
It is important to note that the experiments implementing the mixed Mott insulator in Bose-Fermi mixtures used $^{173}$Yb  and $^{174}$Yb atoms, where the former is a fermion with a spin of $5/2$, i.e., it has six internal degrees of freedom~\cite{Sugawa-NP11}. It is clear that the experimental setting is more complex than that considered in previous theoretical studies. This inspired us to go a step further and consider more degrees of freedom for the fermions. In addition, the physics of SU($N>2$) fermionic systems is very rich and a subject of current interest~\cite{Cazalilla-RPP14,Taie-NP22,Pasqualetti-PRL24}, with a plethora of phenomena to be unravelled. Motivated by this, and seeking to get closer to the experiments, we consider a mixture of scalar bosons and SU($3$) fermions in one dimension. As expected, new physics emerges when $N>2$. Namely, four insulator states emerge, two of them being flavor-selective, which exhibit a three site period unit cell. We observe that bosons can couple with one- or two-flavor fermions to satisfy  a commensurability relation, establishing an insulator state, leaving the others flavors itinerant. Moreover, in addition to the well known mixed Mott insulator, another one that involves all the fermions and bosons arises.\par 
The article is organized as follows. In Sec. ~\ref{sec:model} we introduce the Bose-Fermi Hamiltonian that describes our mixtures, and set up the quantities and definitions used throughout the paper. Our results appear in Sec. ~\ref{sec:results}. Finally, the main findings are summarized in Sec.~\ref{sec:conclusions}.
\section{Bose-Fermi-Hubbard Model}\label{sec:model}

The description of a mixture of an $N$-flavor fermionic gas with scalar bosons in one-dimension requires treating bosons and fermions separately through the Hubbard model, and connecting them through a local interaction term. This leads to the Bose-Fermi-Hubbard model:
\begin{equation}\label{BFHamil}
\begin{split}
\hat{H}=&-t_B\sum_{i} \left( \hat{b}^\dagger_i \hat{b}_{i+1} + \text{H.c.} \right) + \frac{U_{BB}}{2} \sum_i \hat{n}^B_i \left( \hat{n}^B_i - 1 \right) \\
&+ U_{BF}\sum_{i, \alpha} \hat{n}^B_i \hat{n}^F_{ \alpha,i}
-t_F\sum_{i, \alpha} \left( \hat{f}^\dagger_{\alpha, i} \hat{f}_{\alpha, i+1} + \text{H.c.} \right) \\
&+ \frac{U_{FF}}{2} \sum_{i,\alpha \neq \alpha'} 
\hat{f}^\dagger_{\alpha, i}\hat{f}^\dagger_{\alpha', i}\hat{f}_{\alpha', i}\hat{f}_{\alpha, i}.
\end{split}
\end{equation}
Here, the operator $\hat{f}^\dagger_{\alpha, i}$ ($\hat{f}_{\alpha, i}$) creates (annihilates) a fermion of flavor $\alpha=1,\ldots,N$ on site $i$ of a lattice of size $L$. The number operator for the $\alpha$-fermion is $\hat{n}^F_{\alpha,i}=\hat{f}^\dagger_{\alpha, i} \hat{f}_{\alpha, i}$; therefore, the total fermionic number operator is $\hat{n}^F_{i}=\sum_{\alpha}\hat{n}^F_{\alpha,i}$. Given $N^F_{\alpha}$ fermions of flavor $\alpha$, we define their density as $\rho_{\alpha}^F=N^F_{\alpha}/L$ and the total density of fermions is $\rho^F=\sum_{\alpha}\rho_{\alpha}^F$, which varies in the interval $[0, N]$. Also, the operator $\hat{b}_i^\dagger$ ($\hat{b}_i$) creates (annihilates) a boson on site $i$, and $\hat{n}_i^B=\hat{b}_i^\dagger\hat{b}_i$ denotes the boson number operator. We define the bosonic density as $\rho_B=N_B/L$, where  $N_B$ is the number of bosons. The local interaction between fermions (bosons) is quantified by the parameter $U_{FF}$ ($U_{BB}$), whereas $U_{BF}$ measures the boson-fermion repulsion. The hopping amplitude between next neighbor sites for fermions and bosons are $t_F$ and $t_B$, respectively. Taking into account that it is common to confine isotopes of the same kind of atoms in cold-atom setups~\cite{Ikemachi-JPB17,YTakasu-JPSJ09}, we consider $t_F=t_B=1$, which will fix our energy scale.\par
The Bose-Fermi-Hubbard Hamiltonian expressed in Eq.~\eqref{BFHamil} does not have an exact solution. Thus, we resort to a numerical calculation to obtain its ground state. For simplicity, we choose the hard-core approximation for bosons. Therefore, each site can be occupied by one boson at most; the bosonic density varies in the interval $[0, 1]$, and there is no interaction term between bosons. The previous approach establishes that both the local number of bosons and $\alpha$-fermion are either 0 or 1. Therefore, the local basis increases as $2^{N+1}$, and the states are given by all of the combinations $|n_B\rangle |n^F_{1}\rangle \cdots |n^F_{N}\rangle$. To obtain the ground-state energy $E(N_B, N^F_1,\ldots, N^F_N)$ as a function of the number of carriers and the interaction parameters, we use the density matrix renormalization group (DMRG) algorithm implemented in the TeNPy library~\cite{Hauschild-SP18}. In our simulations, we consider a maximum bond dimension of $\chi=6400$ and established a criterion of convergence of  $10^{-5}$ and $10^{-3}$ in energy and entropy, respectively. We use open boundary conditions unless otherwise stated.\par
Finally, important quantities calculated in the paper are the fermionic structure factor 
\begin{equation} \label{struct_factor}
 N^F(k)=\frac{1}{L}\sum_{l,m}e^{ik(l-m)}\Bigl( \langle \hat{n}^F_{l}\hat{n}^F_{m}\rangle - \langle \hat{n}^F_{l}\rangle \langle \hat{n}^F_{m}\rangle \Bigr),
\end{equation} 
and the momentum distribution function of $\alpha$-fermions
\begin{equation}\label{distrib-momen}
 n_{\alpha}(k)=\frac{1}{L}\sum_{l,m} \langle \hat{f}^\dagger_{\alpha, l} \hat{f}_{\alpha, m}  \rangle e^{ik(l-m)}.
\end{equation} 
These quantities will be crucial for locating the quantum critical points of the model, and for determining the nature of the corresponding ground states.
\begin{figure}[t!]
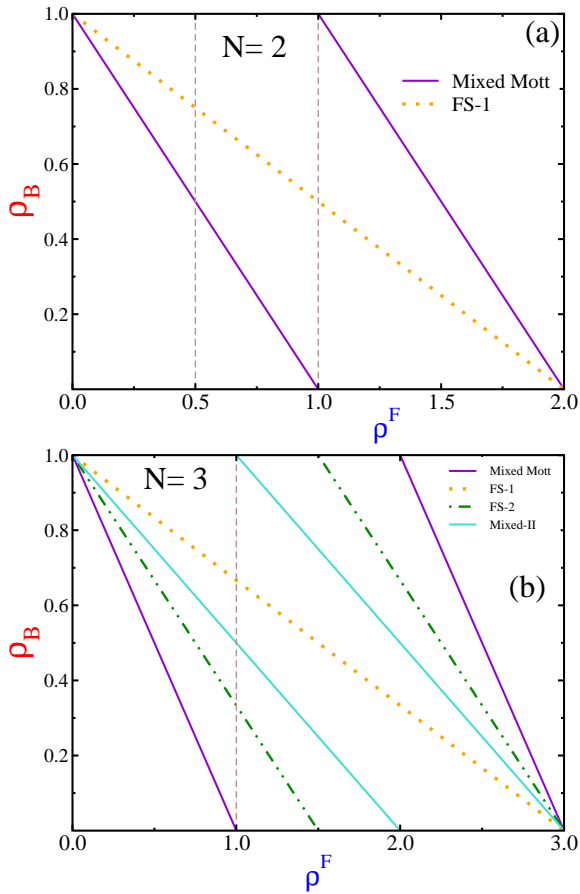

\begin{minipage}{18pc}
\includegraphics[width=18pc]{Fig1a.eps}
\end{minipage}
\hspace{7pc}%
\begin{minipage}{18pc}
\includegraphics[width=18pc]{Fig1b.eps}
\caption{\label{fig1} Possible insulator phases of a balanced mixture of SU($N$) fermions and scalar bosons in one dimension. For rational values of the  bosonic density ($\rho_B$) and the fermionic density ($\rho^F$), we draw the corresponding lines of the relations that determine each insulator for $N=2$ (a) and $3$ (b). Outside the lines, the points $(\rho^F,\rho_B)$ correspond to a gapless phase. To identify the insulating states, we fix $N$, then choose a value for $\rho^F$ (by drawing a brown dashed vertical line), and we then continuously increase the number of bosons from zero. Every time the vertical line touches an oblique line, the indicated insulator arises (for suitable values of the interaction parameters). Otherwise, the states along the vertical line will be gapless. The notation FS-$\gamma$ indicates a flavor-selective insulator where the bosons couple to $\gamma=1$ or $2$ fermionic flavors to fulfill the commensurability relation with the lattice. Mixed Mott refers to mixed Mott insulator, while Mixed-II corresponds to the new non flavor-selective phase found for $N=3$ (see Section~\ref{sec:imbalance}).}
\end{minipage}\hspace{2pc}%
\end{figure}
\section{Results}\label{sec:results}

A degenerate gas composed of only fermions or bosons already features very interesting physics. The phenomenology that emerges when combining them is vast and intriguing. Here, we consider scalar bosons and explore the insulating phases that flourish as the number of internal fermionic degrees of freedom increases. Combining polarized fermions ($N=1$) and bosons opened the possibility of a new insulator state ruled by the commensurability condition and the repulsive coupling. This state is known as mixed Mott insulator, and fulfills the relation $\rho_B + \rho^F=1$ ~\cite{Zujev-PRA08,Sugawa-NP11}. Hence, the possible ground states can be sumed up in a graph of $\rho_B$ versus $\rho^F$, where the mixed Mott insulator is represented by a decreasing straight line with slope -1. From now on, we only consider rational values of the bosonic ($\rho_B$) and fermionic ($\rho^F$) densities. When $\rho_B=0$, the mixed Mott insulator correspond to the trivial fermion band insulator, with a fermion in every lattice site.\par
By introducing one more degree of freedom for the fermions, we obtain a mixture of two-flavor fermions ($N=2$) and scalar bosons, whose $\rho_B-\rho^F$ phase diagram is shown in Fig. ~\ref{fig1} (a). The oblique colored lines represent the density coordinates for which the gapped phases arise for suitable values of the couplings, while the other points correspond to gapless states. As expected, the mixed Mott insulator appears (continuous violet line for $0<\rho^F<1$), as it will do for any value of $N$, due to the fact that in this state, the total number of carriers (bosons + fermions) will be commensurate with the lattice in a mixture of interacting carriers. Hamiltonian ~\eqref{BFHamil} satisfies the particle-hole symmetry; thus, another mixed Mott line is seen for $\rho^F>1$ (continuous violet line for $1<\rho^F<2$). As shown by us in a previous work, another gapped phase emerges, namely the flavor-selective insulator, for which one kind of fermions is in a gapless state, while the other is in a gapped one and satisfies a commensurability relation with the bosons~\cite{Avella-PRA19,Avella-PRA20,GuerreroS-PRA21,JSV-RACCFYN22}. For a balanced mixture ($\rho_{1}^F=\rho_{2}^F$), the flavor-selective insulator is non-magnetic and satisfies the relation $\rho_B + \rho_{\alpha}^F=1$ ($\alpha=1$ or $2$), which corresponds to the orange dotted curve in Fig. ~\ref{fig1} (a). From now on, we will denote this state as FS-1, which means that in this particular flavor-selective state the bosons are coupled to only one kind of fermions and both fulfill a commensurability relation with the lattice. Under a spin-population imbalance ($\rho_{1}^F\neq \rho_{2}^F$) the mixed Mott insulator remains unaltered, whereas the non-magnetic FS-1 splits into two ferromagnetic flavor-selective insulators, which fulfill $\rho_B + \rho_{1}^F=1$ and $\rho_B + \rho_{2}^F=1$. 
This was demonstrated for $\rho^F= 0.5$ in Fig. 2 of reference ~\cite{GuerreroS-PRA21}, and is shown in Fig.~\ref{fig1}(dashed brown vertical line). As a test of the consistency of the results presented in Fig. ~\ref{fig1} (a), we assess the limit $\rho_B= 0$ (without bosons). We recover the gapped phases of the SU($2$) Fermi-Hubbard model at $\rho^F= 1$ (antiferromagnetic Mott insulator) and $\rho^F= 2$ (band insulator). In summary, for a balanced mixture with $N=2$, there are two gapped states: the mixed Mott insulator and the non-magnetic FS-1 one.\par 
\begin{figure}[t]
\centering
\includegraphics[width=20pc]{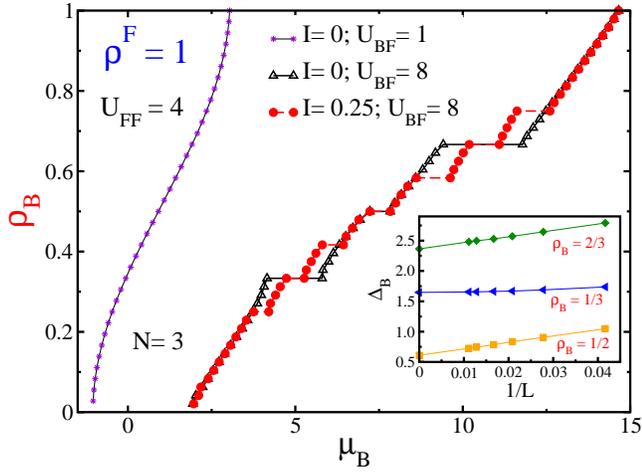}
\caption{Bosonic density $\rho_B$ as a function of bosonic chemical potential $\mu_B$, calculated in the thermodynamic limit, for a mixture of scalar bosons and three-flavor fermions ($N=3$). The fixed parameters are the fermionic density $\rho^F=1$ and repulsive fermion-fermion interaction $U_{FF}=4$. The stars (violet line) and the open triangles (black line) correspond to a balanced ($I=0$) mixture for boson-fermion coupling $U_{BF}=1$ and $U_{BF}=8$, respectively. Furthermore, a curve for an imbalance mixture ($I=0.25$) is shown for $U_{BF}=8$ (closed circles, red line). Inset: Charge gap $\Delta_B$ as a function of the system size for the bosonic densities that correspond to the plateaus of the $I=0$, $U_{\text{BF}}=8$ curve. The diamonds on the $y$ axis correspond to the extrapolations to the thermodynamic limit. The lines are visual guides.}
\label{fig2}
\end{figure}
\subsection{Balanced flavor populations}\label{sec:balanced}
From this point on, we will discuss the physics of mixtures of scalar bosons and three-flavor ($N=3$) fermions in one dimension. Figure~\ref{fig1} (b) shows the possible gapped states of a balanced mixture, i.e., when there is an equal number of fermions for each flavor. This scenario is characterized by $I=0$, where $I$ is the imbalance parameter to be defined later. In addition to the mixed Mott lines, we observe more lines than in the $N=2$ system, indicating a richer physics here. To concentrate in the unknown states, we will fix the fermionic density at $\rho^F= 1$ (dashed brown vertical line) and the fermion-fermion repulsion $U_{FF}=4$. Next, we will increase the number of bosons from zero for two different values of the boson-fermion coupling, $U_{BF}=1$ and $8$. As shown in Fig.~\ref{fig2}, for the former value of the boson-fermion repulsion, the thermodynamic-limit value of the bosonic chemical potential
\begin{multline}
 \mu_B = E(N_B, N^F_1,N^F_2, N^F_3) - \\
E(N_B-1, N^F_1,N^F_2, N^F_3)
\end{multline}
varies continuously as the number of bosons increases (stars, violet line). This result indicates that the ground state will be gapless for this set of parameters. To visualize the scenario depicted in Fig.~\ref{fig1} (b), where the insulating states emerge at the intersections between the dashed brown vertical line and the other lines, it is necessary to increase the coupling between bosons and fermions. Hence, there will be a critical value of $U^{*}_{BF}$ for the emergence of the unknown insulating states, in a similar way to the cases $N=1$~\cite{Zujev-PRA08} and $N=2$ ~\cite{Avella-PRA19,Avella-PRA20,GuerreroS-PRA21,JSV-RACCFYN22}. For $U_{BF}= 8$, the $\rho_B$-$\mu_B$ curve breaks its continuous behavior at the densities $\rho_B=1/3$, $1/2$ and $2/3$ (open triangles, black line). This coincides with the intersections shown in Fig.~\ref{fig1} (b) between the dashed brown vertical line and the other three lines, indicating that three gapped phases arise. In the inset of Fig.~\ref{fig2}, we observe the evolution of the bosonic charge gap,  
\begin{align}
\Delta_B =& E(N_B+1, N^F_1,N^F_2, N^F_3)+\\
& E(N_B-1, N^F_1,N^F_2, N^F_3)-2E(N_B, N^F_1,N^F_2, N^F_3),\notag
\end{align}
with the inverse of the lattice size. In the thermodynamic limit, the gap at bosonic densities $\rho_B=1/3$, $1/2$ and $2/3$ is finite and takes the values $\Delta_B^{\rho_B=1/3} = 1.65$, $\Delta_B^{\rho_B=1/2} = 0.61$, and $\Delta_B^{\rho_B=2/3} = 2.36$, respectively. 
Therefore, the unveiled insulating states are physical and not a numerical artifact.\par
\begin{figure}[t!]
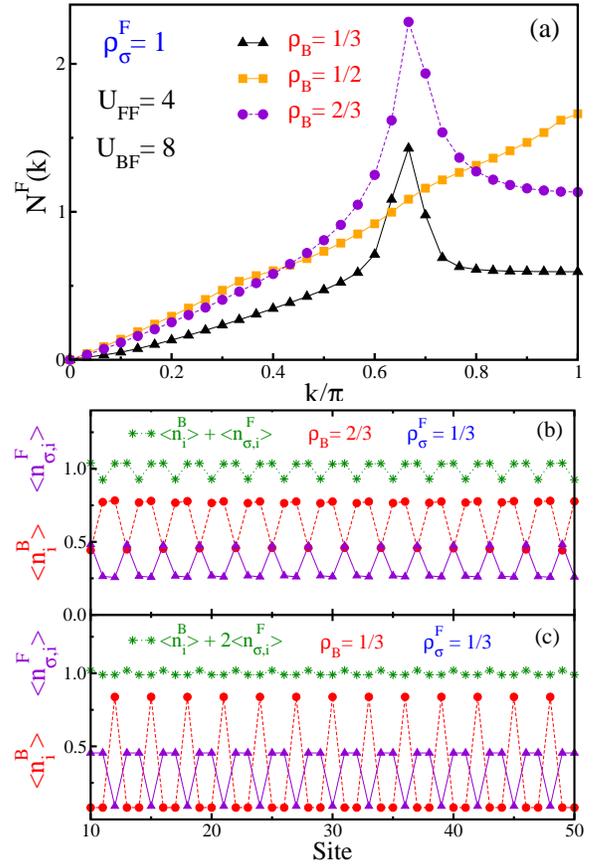

\begin{minipage}{18pc}
\includegraphics[width=18pc]{Fig3a.eps}
\end{minipage}
\hspace{7pc}%
\begin{minipage}{18pc}
\includegraphics[width=18pc]{Fig3b.eps}
\caption{\label{fig3}Fermionic structure factors and density profiles for a balanced mixture of scalar bosons and three-flavor fermions ($N=3$). The calculations were done at the insulator states characterized by the bosonic densities $\rho_B=1/3$, $1/2$ and $2/3$. The global fermionic density is $\rho^F=1$, the system size is $L=60$, and the interaction parameters are 
$U_{FF}=4$ and $U_{BF}=8$. (a) Fermionic structure factors. In (b) and (c), we show the density profiles at the insulator states with $\rho_B=2/3$ and $1/3$, respectively. We use periodic boundary conditions to obtain (b) and (c).
The lines are visual guides.}
\end{minipage}\hspace{2pc}%
\end{figure}
Mixing fermions with three internal degrees of freedom and scalar bosons generates four gapped states, namely the mixed Mott insulator and three whose nature will be determined. Although we do not show explicitly the mixed Mott insulator in this paper, it was observed for global fermionic densities $\rho^F<1$, and it fulfills the relation $\rho_B + \rho^F=1$, which correspond to the violet continuous line in Fig.~\ref{fig1} (b). Open triangles (black curve) in Fig.~\ref{fig2} correspond to a balanced mixture with $\rho^F=1$, for which $\rho_{\alpha}^F=1/3$; this suggests possible links between bosons and some flavors of fermions to establish insulators that fulfill commensurability relations with the lattice. To gain more intuition on these states, we plot in Fig.~\ref{fig3}(a) the total fermionic density structure factor Eq.~\eqref{struct_factor}
at the bosonic densities $\rho_B= 1/3$, $1/2$ and $2/3$. It is clear that the insulator at $\rho_B=1/2$ differs from the others, as the latter have a peak at $k=2\pi/3$ whereas the former grows monotonously. Note that the same information can be obtained from the bosonic structure factor (not shown). A strong peak in the structure factor evidences certain periodicity in the carrier  distribution, which in this case indicates the predominance of an unit cell composed of three lattice sites. However, we cannot differentiate the insulators at $\rho_B=1/3$ and $2/3$ from this figure. To observe the carrier distribution, we plot the density profiles of bosons and fermions at the bosonic densities $\rho_B=2/3$ and $1/3$ in Figs.~\ref{fig3}(b) and (c), respectively. As the number of each flavor fermion is the same (balanced mixture $I=0$), which implies that $\rho^F_{\alpha}=1/3$, we display only the density profile of one flavor fermion. At $\rho_B=2/3$, we observe that the expectation values of the number of bosons and fermions oscillate off-phase across the lattice, with a unit cell characterized by one lowly-occupied and two highly-occupied sites for bosons, and the opposite for fermions. In each unit cell, there are two bosons and three fermions (one of each flavor). Adding up locally $\langle n^B_i\rangle$ and $\langle n^F_{\alpha,i}\rangle$ we obtain values that oscillates around 1 across the lattice, with a three-sites unit cell. The above suggests the commnensurability relation $\rho_B + \rho^F_{\alpha}=1$ between the bosons and one-flavor fermion, which is corroborated by adding up the populations throughout the lattice, i.e., the sum of the number of bosons and one-flavor fermion coincides with the lattice size. Thus, the gapped state at $\rho_B=2/3$ is an flavor-selective insulator where the bosons are coupled to one-flavor fermions to establish an insulator, while the other two-flavor fermions remain itinerant. This result generalizes the findings for the $N=2$ case~\cite{GuerreroS-PRA21}. This flavor-selective state, referred to as FS-1 and represented by the orange dotted line in Fig.~\ref{fig1}(b), fulfills the relation $\rho_B + \tfrac{1}{3}\rho^F=1$.\par
In Fig.~\ref{fig3}(c), the $\rho_B=1/3$ density profiles of bosons and one-flavor fermions exhibit a unit cell with two lowly-occupied sites and one highly-occupied one for bosons, and the opposite for fermions. At each unit cell there are one boson and three fermions (one of each flavor). Looking at the local values of $\langle n^B_i\rangle$ and $\langle n^F_{\alpha,i}\rangle$, it is clear that their sum is not close to 1; therefore, the gapped state at $\rho_B=1/3$ is different from that with $\rho_B=2/3$. Namely, we found that adding $\langle n^B_i\rangle$ and two times $\langle n^F_{\alpha,i}\rangle$ we get values closer to 1 throughout the lattice with an underlying oscillation of a period of three sites. Hence, our results suggest that gapped state at $\rho_B=1/3$ is characterized by the coupling of bosons and two-flavor fermions to establish an insulator, while the other flavor fermion is in a gapless state. This reveals a new kind of flavor-selective insulator, satisfying the relation $\rho_B + 2\rho^F_{\alpha}=1$. The above relation is also written in terms of the global fermionic density as $\rho_B + \tfrac{2}{3}\rho^F=1$, which is represented by the green dashed dotted line in Fig.~\ref{fig1} (b). This shows the evolution of the flavor-selective state referred to as FS-2 due to the coupling of bosons with two flavors of fermions.\par
In brief, for a mixture of three-flavor fermions and scalar bosons, we find the emergence of two flavor-selective insulators and another one with no particular ordered carrier distribution.\par 
%
%
Figure~\ref{fig2} shows that for a given global fermionic density ($\rho^F$) and a fermionic intraspecies interaction ($U_{FF}$), the ground state of a mixture of three-flavor fermions and scalar bosons will be determined by the bosonic density ($\rho_B$) and the particular value of the intersepecies interaction $U_{BF}$. These observations open up the possibility of a non-trivial dependence of the phase diagram with the latter parameter.\par
\begin{figure}[t]
\centering
\includegraphics[width=21pc]{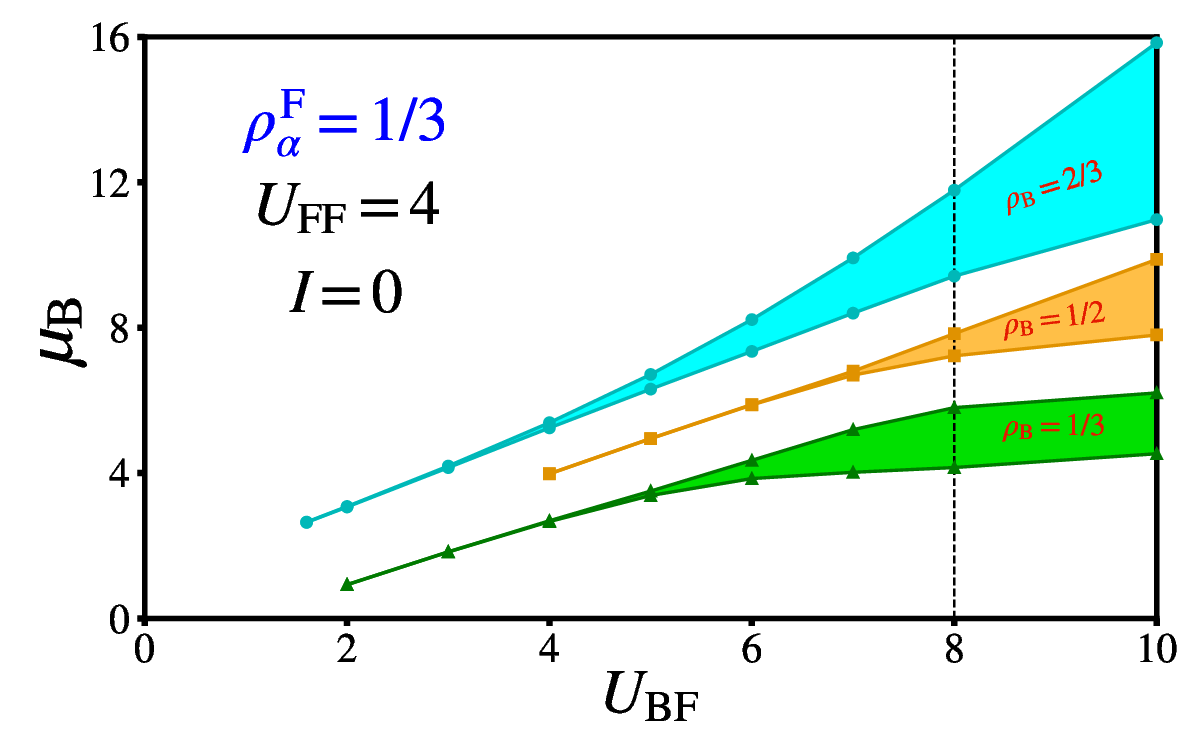}
\caption{Bosonic chemical potential $\mu_B$  as a function of the boson-fermion interaction $U_{BF}$ for a balanced mixture of scalar bosons and three-flavor fermions ($N=3$), forming the phase diagram of the system. The fixed parameters are the fermionic density $\rho^F=1$ and the repulsive fermion-fermion interaction $U_{FF}=4$. The vertical dashed line corresponds to the black curve in Fig.~\ref{fig2}. The lines that join the extrapolated values of the chemical potential in the thermodynamic limit are guides to the eyes.}
\label{fig7}
\end{figure}
The emergence and growth of the insulator lobes as a function of the interspecies interaction, in a balanced Bose-Fermi mixture with a global fermionic density $\rho^F=1$ and $U_{FF}=4$, are shown in Fig.~\ref{fig7}. The vertical dashed line at $U_{BF}=8$ corresponds to the already discussed black curve in Fig.~\ref{fig2}. The length of the plateaus is in correspondence with the width of the insulator lobes shown in Fig.~\ref{fig7}. Therefore the white area in the figure represents a gapless ground state (superfluid), while the colorful areas are related to gapped ground states. It is clear from the figure that all the insulator lobes grow with the boson-fermion interaction; however, their increase rate varies for each insulator. Similarly, the interaction from which each insulator lobe emerges will be different. Namely, the critical interspecies interaction for the insulator lobe with bosonic densities $\rho_B=1/3$, $\rho_B=1/2$, and $\rho_B=2/3$ are $U^{*}_{BF}=2.0 \pm 0.3$,  $U^{*}_{BF}=4.0 \pm 0.4$, and $U^{*}_{BF}=1.6 \pm 0.3$, respectively.\par 
\subsection{Imbalanced flavor populations}\label{sec:imbalance}
We assert that some of the insulating states revealed in this study are flavor dependent; however, additional information is necessary in order to fully characterize the nature of the such states. To obtain this knowledge, we need to break the SU($3$) symmetry, which was done recently in cold-atom experiments with Raman spectroscopy~\cite{Tusi-NP22}. Here, we break the SU($3$) symmetry by considering different flavor populations, and we quantify the population imbalance through the parameter 
\begin{align}
I= (N^F_1-N^F_3)/(N^F_1+N^F_3)
\end{align}
for the $N= 3$ case; therefore a balanced mixture means that $I= 0$.\par 
To evidence the flavor-selective characteristic of the gapped states at 
$\rho_B=1/3$ and $2/3$, we consider an imbalanced mixture with $I=0.25$, which corresponds to $\rho^F_{1}=5/12$, $\rho^F_{2}=1/3$ and $\rho^F_{3}=1/4$, keeping fixed the global fermionic density ($\rho^F=1$) as well as the interaction parameters used in our previous discussions. The curve of red circles in Fig.~\ref{fig2} displays a rich scenario with several plateaus. The imbalanced mixture has seven gapped states at the bosonic densities $\rho_B=1/4$, $\rho_B=1/3$, $\rho_B=5/12$, $\rho_B=1/2$, $\rho_B=7/12$, $\rho_B=2/3$, and $\rho_B=3/4$, each of which has a finite charge gap in the thermodynamic limit equal to $\Delta_B^{\rho_B=1/4} = 0.45$, $\Delta_B^{\rho_B=1/3} = 0.53$, $\Delta_B^{\rho_B=5/12} = 0.63$, $\Delta_B^{\rho_B=1/2} = 0.61$, $\Delta_B^{\rho_B=7/12} = 1.01$, $\Delta_B^{\rho_B=2/3} = 0.90$, $\Delta_B^{\rho_B=3/4} = 0.98$, respectively. Note that around $\rho_B=1/2$ the balanced (open black triangle) and imbalance (full red circles) curves in Fig.~\ref{fig2} coincide, implying that the imbalance does not affect the gapped state at $\rho_B=1/2$.\par 
\begin{figure}[t!]
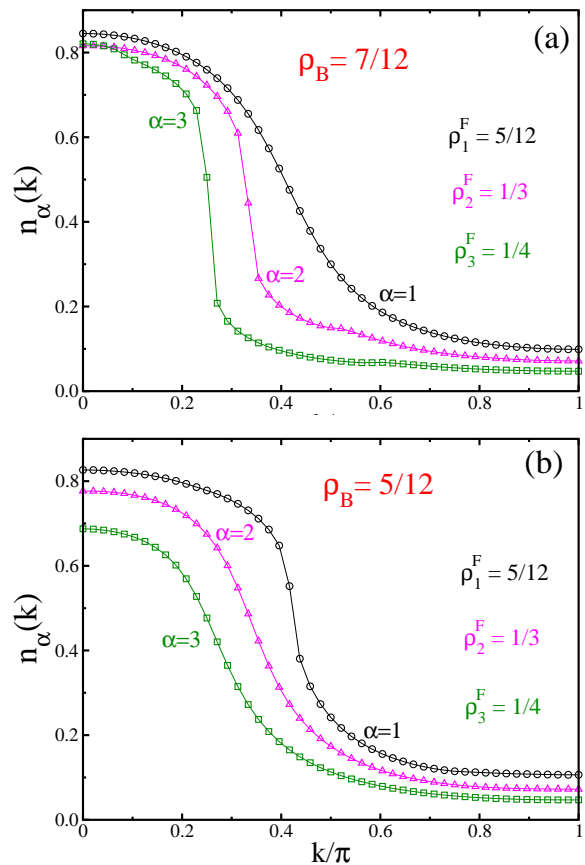

\begin{minipage}{18pc}
\includegraphics[width=18pc]{Fig4a.eps}
\end{minipage}
\hspace{7pc}%
\begin{minipage}{18pc}
\includegraphics[width=18pc]{Fig4b.eps}
\caption{\label{fig4} Momentum distribution function for each $\alpha$-fermion at the insulator states with bosonic densities $\rho_B=7/12$ (a) and $\rho_B=5/12$ (b) for an imbalanced mixture of scalar bosons and three-flavor fermions ($N=3$). Here, we fixed the lattice size to $L=60$, the global fermionic density to $\rho^F=1$, the imbalance to $I=0.25$, the repulsive fermion-fermion interaction $U_{FF}=4$, and the boson-fermion coupling to $U_{BF}=8$. The lines are visual guides.}
\end{minipage}\hspace{2pc}%
\end{figure}
Our particular rule to generate the imbalance keeps the fermion $\alpha =2$ density ($\rho^F_{2}=1/3$) fixed, allowing to see clearly that the imbalance splits the non-magnetic insulators found in the balanced case at $\rho_B=1/3$ and $\rho_B=2/3$. The latter non-magnetic insulator splits in three gapped states with bosonic densities $\rho_B=7/12$, $\rho_B=2/3$, and $\rho_B=3/4$, which means that in each case the bosons couple with one flavor fermion to establish an insulator, while the others remains in gapless states. For instance, at 
$\rho_B=7/12$, the bosons couple with the fermions $\alpha =1$ ($\rho^F_{1}=5/12$), fulfilling the commensurability relation $\rho_B + \rho^F_{1}=1$. To reinforce the above statement, we can look at the momentum distribution function of fermions defined by Eq.~\eqref{distrib-momen}. It is known that for non-interacting fermions, the momentum distribution function is given by a sharp step (rectangular) function around the Fermi wave vector $k_F$, indicating that the states above (below) $k_F$ are empty (full). Of course, as the fermions are free, the system is gapless and an algebraic decay of the correlation functions is expected. In a gapless state and in presence of interactions a step-like feature is still present in the momentum distribution function; however, the size of the jump is lower than 1, which quantifies the effective interactions experienced by the fermions. On the contrary, within a Mott insulating phase the momentum distribution function evolves continuously and smoothly around $k_F$~\cite{Rizzi-PRA08,Manmana-PRA11,Schonmeier-Kromer-PRB23}. In Fig.~\ref{fig4}(a), we show the momentum distribution function $n^{F}_{\alpha}(k)$ for each flavor of fermions at the gapped state with bosonic density $\rho_B=7/12$. Clearly, the momentum distribution function changes around $\rho^F_{\alpha}$ for any fermionic flavor, but the most important feature is how this change happens. We observe a step-like behavior in the momentum distribution function curves for fermions $\alpha =2$  and $\alpha =3$, which indicates that these fermions are in gapless phases, whereas $n^{F}_{1}(k)$ passes smoothly and continuously through $\rho^F_{1}$, a behavior related with a Mott insulator state. The previous analysis allow us to affirm that at the gapped state with bosonic density $\rho_B=7/12$, the fermions $\alpha =2$ and $\alpha =3$ remain in gapless states while the bosons and fermions $\alpha =1$ are coupled, establishing an insulator state that satisfies a commensurability relation with the lattice, given by $\rho_B + \rho^F_{1}=1$. Of course similar arguments can be applied to the gapped states at $\rho_B=2/3$, and $\rho_B=3/4$, where the bosons couple with the fermions $\alpha =2$ and $\alpha =3$, respectively.\par

\begin{figure}[t!]
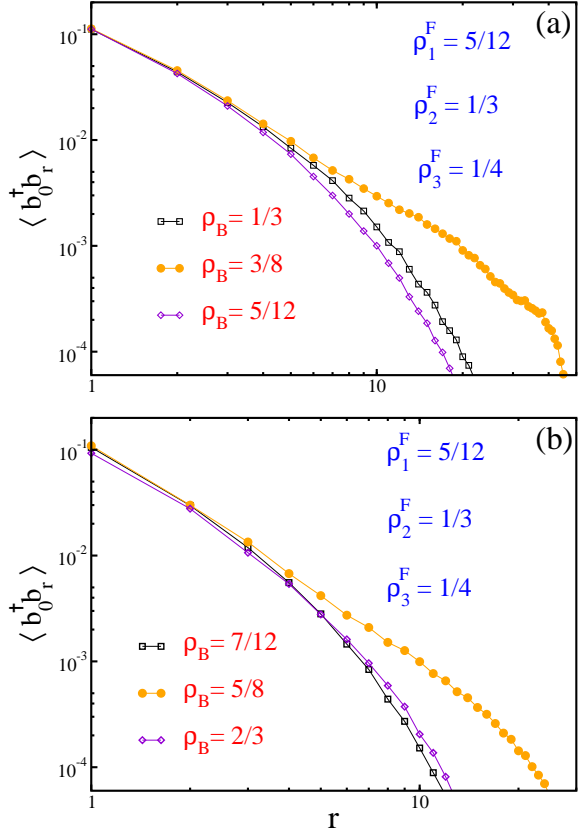

\begin{minipage}{18pc}
\includegraphics[width=18pc]{Fig5a.eps}
\end{minipage}
\hspace{7pc}%
\begin{minipage}{18pc}
\includegraphics[width=18pc]{Fig5b.eps}
\caption{\label{fig5} Bosonic correlations $\langle b^{\dagger}_0b_r \rangle$ as a function of  distance $r$ for an imbalance mixture of scalar bosons and three-flavor fermions ($N=3$). The fixed parameters are the lattice size $L=96$, the global fermionic density $\rho^F=1$, the imbalance $I=0.25$, the repulsive fermion-fermion interaction $U_{FF}=4$, and the boson-fermion coupling $U_{BF}=8$. The points are DMRG results, while the lines are guide for the eyes. The bosonic correlations for a  gapless state ($\rho_B=3/8$ and $\rho_B=5/8$) follow the relation $\langle b^{\dagger}_0b_r \rangle=A\lvert r \rvert^{-B}$, whereas for gapped states it is $\langle b^{\dagger}_0b_r \rangle=Ce^{-Dr}$, where $A$, $B$, $C$ and $D$ are fitting parameters. The particular values of the above parameters for each bosonic density in this figure are: $C^{\rho_{B}=1/3}=0.035$, $D^{\rho_{B}=1/3}=0.287$, $R^2_{\rho_{B}=1/3}=0.993$; $A^{\rho_{B}=3/8}=0.265$, $B^{\rho_{B}=3/8}=2.006$, $R^2_{\rho_{B}=3/8}=0.932$; $C^{\rho_{B}=5/12}=0.034$, $D^{\rho_{B}=5/12}=0.328$, $R^2_{\rho_{B}=5/12}=0.994$; $C^{\rho_{B}=7/12}=0.032$, $D^{\rho_{B}=7/12}=0.508$, $R^2_{\rho_{B}=7/12}=0.997$; $A^{\rho_{B}=5/8}=0.309$, $B^{\rho_{B}=5/8}=2.670$, $R^2_{\rho_{B}=5/8}=0.968$; $C^{\rho_{B}=2/3}=0.029$, $D^{\rho_{B}=2/3}=0.472$, $R^2_{\rho_{B}=2/3}=0.997$. Here, $R^2$ is the square of the correlation coefficient. }
\end{minipage}\hspace{2pc}%
\end{figure}
\begin{figure}[t]
\centering
\includegraphics[width=19pc]{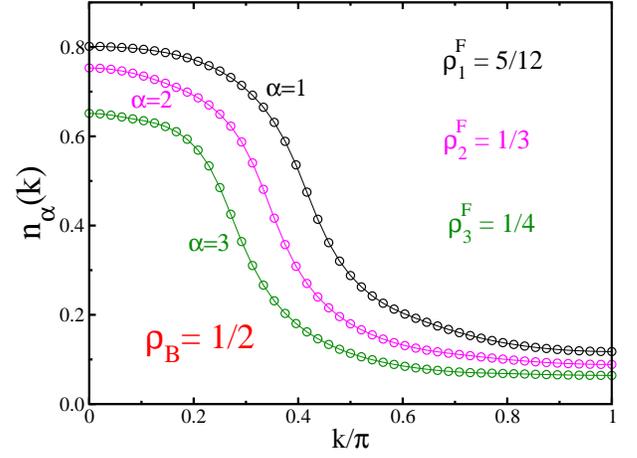}
\caption{Momentum distribution function for each $\alpha$-fermion at the insulator state with bosonic densities $\rho_B=1/2$ for an imbalanced mixture of scalar bosons and three-flavor fermions ($N=3$). We fixed the lattice size to $L=96$, the global fermionic density to $\rho^F=1$, the imbalance to $I=0.25$, the repulsive fermion-fermion interaction to $U_{FF}=4$, and the boson-fermion coupling to $U_{BF}=8$. The lines are visual guides.}
\label{fig6}
\end{figure}
%
Our previous intuition on the non-magnetic gapped state at $\rho_B=1/3$ (open black triangles in Fig.~\ref{fig2}) suggests that in this case, the bosons couple with two kinds of fermions. Since we have three flavor fermions, we can form three possible combinations of boson-fermion arrangements. Thus, the non-magnetic plateau at $\rho_B=1/3$ must split into three gapped states located at $\rho_B=1/4$, $\rho_B=1/3$, and $\rho_B=5/12$. Looking closely at the latter gapped state, we see that the only way to fulfill a commensurability relation with the lattice is to couple the bosons ($\rho_B=5/12$), the fermions $\alpha =2$ ($\rho^F_{2}=1/3$) and the fermions $\alpha =3$ ($\rho^F_{3}=1/4$), obtaining that $\rho_B+\rho^F_{2} +\rho^F_{3} =1$. Figure~\ref{fig4}(b) shows the momentum distribution function for fermions at the gapped state with bosonic density $\rho_B=5/12$. In contrast to Fig.~\ref{fig4}(a), the momentum distribution function for fermions $\alpha =1$ exhibits a step-like behavior around $\rho^F_{1}$, while $n^F_{2}(k)$ and $n^F_{3}(k)$ evolve continuously and smoothly. Hence, we can affirm that fermions $\alpha =2$ and $\alpha =3$ are in an insulator phase, whereas fermions $\alpha =1$ are in a gapless one. We conclude that the gapped state at $\rho_B=5/12$ is a flavor-selective insulator where the bosons combine with fermions $\alpha =2$ and $\alpha =3$ to establish an insulator. Similar argumentation allow us to assert that the gapped states located at $\rho_B=1/4$ and $\rho_B=1/3$ are flavor-selective insulators that fulfill the relations $\rho_B+\rho^F_{1} +\rho^F_{2} =1$ and $\rho_B+\rho^F_{1} +\rho^F_{3} =1$, respectively.\par 
At this point, it is clear that for a mixture of scalar bosons and three-flavor fermions, the non-magnetic flavor-selective insulators located at $\rho_B=1/3$ and $\rho_B=2/3$ are each divided into three ferromagnetic flavor-selective insulators due to the imbalance. Figure~\ref{fig2} suggests that the latter flavor-selective insulators are separated from each other by gapless regions, commonly expected to be superfluids; however, additional confirmation is necessary to characterize these states. With this in mind, we calculated the bosonic correlation function $\langle b^{\dagger}_0b_r \rangle$, from site $j=0$ in the center of the lattice, as a function of the distance $r$ for several bosonic densities. Considering the bosonic densities $\rho_B=1/3$ and $\rho_B=5/12$, we see in Fig.~\ref{fig5}(a) that $\langle b^{\dagger}_0b_r \rangle$ decreases monotonously and can be fitted to a exponential function. This indicates that the ground state is an insulator for these densities, as already evidenced by characterizing other quantities. Also a exponential decay of the bosonic  correlation function is clearly observed in Fig.~\ref{fig5}(b) for the bosonic densities $\rho_B=7/12$ and $\rho_B=2/3$, as expected. However, for $\rho_B=3/8$ in Fig.~\ref{fig5}(a), which is located between $\rho_B=1/3$ and $\rho_B=5/12$, $\langle b^{\dagger}_0b_r \rangle$ decreases in a different way. A similar scenario is seen for $\rho_B=5/8$ in Fig.~\ref{fig5}(b), which is located between $\rho_B=7/12$ and $\rho_B=2/3$. The ground state for the bosonic densities $\rho_B=3/8$ and $\rho_B=5/8$ is gapped, and the bosonic correlation function decays following a power-law, i.e.  $\langle b^{\dagger}_0b_r\rangle \sim \lvert r \rvert^{-K^{*}/2}$~\cite{Kuhner-PRB00}. After several fittings considering different ranges of $r$, we obtain values of $K^{*}>2$, which suggest that the ground state is a polaronic Luttinger liquid~\cite{Mathey-PRL04}. Therefore, the insulator states reported here are surrounded by gapless states, i.e., superfluids.\par 
Figure~\ref{fig2} shows that the insulator state with bosonic density $\rho_B=1/2$ remains unaltered after the SU($N$) symmetry is broken, i.e., this gapped state is not flavor-selective, similarly to the mixed Mott insulator. Its structure factor grows continuously and reaches a maximum value at $k=\pi$ (Fig.~\ref{fig3}(a)). Within this insulator state, for every two neighboring sites, there is one boson and two total fermions composed of all flavors. To verify the cooperative character of the fermions at this gapped state, we display in Fig.~\ref{fig6} the momentum distribution function $n^{F}_{\alpha}(k)$ (Eq.~\eqref{distrib-momen}) for each fermionic flavor. Contrasting Figs.~\ref{fig4} and ~\ref{fig6}, we see that in the latter case all fermions feature the same behavior. For any flavor, the momentum distribution function passes continuously through $k_{F,\alpha}$, indicating that all fermions are in an insulator state. Therefore, the emergence of this state depends solely on the global fermionic density ($\rho^F$) and the bosonic density ($\rho_B$), as it also happens for the mixed Mott insulator. Exploring the conditions that lead to the emergence of the gapped state at $\rho_B=1/2$, that we refer to as Mixed-II, we observed that this state survives in the thermodynamic limit, and it emerges for a wide range of densities. We have established that it satisfies the relation $\rho_B + \tfrac{1}{2}\rho^F=1$, which has been tested for several fermionic densities. The above relation corresponds to the turquoise continuous line in Fig.~\ref{fig1}(b); there, we also show the equivalent curve generated by the particle-hole symmetry of the model.\par 
In Fig.~\ref{fig7}, we showed the insulating lobes as a function of the interspecies interaction $U_{BF}$. Under a flavor-population imbalance, the central insulator lobe that corresponds to the bosonic density $\rho_B=1/2$ will remain unaltered. On the other hand, the other flavor-selective lobes will be divided into three lobes, as shown for the specific case of $U_{BF}=8$ (red curve in Fig.~\ref{fig2}). We expect that the lobes arising from the ones at $\rho_B=1/3$ and $\rho_B=2/3$ emerge from different values of $U_{BF}$ for each one.\par
In the absence of bosons ($\rho_B= 0$) Fig.~\ref{fig1} (b) predicts three non-trivial insulator states at the fermionic densities $\rho_F=1$, $3/2$ and $2$. The first two have been widely studied for the SU($3$) Hubbard model and now we know that the gapless-insulator transition takes place at $U^{*}_{FF}\approx1.5$ for $\rho_F=1$, while at the half-filled case, the transition occurs at $U^{*}_{FF}=0$~\cite{Assaraf-PRB99,Buchta-PRB07,Manmana-PRA11}. Our results show that the insulator at $\rho_F=2$ derived from the Mixed-II insulator, which  satisfies the relation $\rho_B + \tfrac{1}{2}\rho^F=1$, exhibits the same characteristics as the insulator at the fermionic density $\rho_F=1$. \par 
\section{Conclusions} \label{sec:conclusions}
We have explored the ground state of a system composed by SU($3$) fermions and scalar bosons, which interact locally and repulsively among each other. Using the density matrix renormalization group method, we identified the emergence of novel insulator states as we sweep throughout the system parameters, such as the bosonic and fermionic densities, and the coupling strengths, establishing several phase diagrams.\par 
It is well known that a mixture of bosons and fermions always exhibits a mixed Mott insulator, where the total number of carriers (bosons + fermions) will be commensurate with the lattice, leading to the relation $\rho_B + \rho^F=1$. In the mixed Mott insulator, all flavors of fermions participate; hence, this state is not flavor-selective and will be present for any number of internal degrees of freedom in bosons or fermions.\par
Taking into account the internal degrees of freedom for fermions allows the emergence of diverse insulator states. In a mixture of scalar bosons and three-flavor fermions, three additional insulators were identified, one of them being non-flavor-selective and satisfying the relation $\rho_B + \tfrac{1}{2}\rho^F=1$, which we labeled Mixed-II.\par 
With or without the SU($3$) symmetry, flavor-selective insulators emerge, in which bosons tie to certain flavors of fermions while the others remain itinerant. These states have a three-site unit cell, where three fermions couple with one or two bosons. When only one kind of fermionic flavor ties to the bosons, this flavor-selective state fulfills  $\rho_B + \rho^F_{\alpha}=1$, as in the $N=2$ case~\cite{Avella-PRA19,Avella-PRA20,GuerreroS-PRA21,JSV-RACCFYN22}. Considering $N>2$ allows to unearth the flavor-selective state that tie bosons and two flavors of fermions, while the remaining flavor remain itinerant. This state satisfies the relation $\rho_B + \rho^F_{\alpha}+\rho^F_{\alpha'}=1$.\par 
Despite the fact that the hard-core boson condition might be strongly constraining (although necessary if the main goal is to unveil the physics of this mixture), we anticipate that relaxing the hard-core condition will lead us to find four gapped states between every two trivial bosonic Mott insulators, similarly to our conclusions for the SU($2$) case~\cite{Avella-PRA20}. \par
Preliminary results for $N=4$ suggest that flavor-selective insulators that couples bosons with one, two or three flavor fermions will appear. However, no other non flavor-selective state is revealed, which suggests that this feature belongs only to an odd number of internal degrees of freedom for the fermions. This insight will be explored in a future research.\par
\section*{Acknowledgments}
J. S.-V. thanks C. Quimbay for asking interesting questions that inspire this work. J. S.-V. report financial support was provided by ``\textit{Fundación Para la Promoción de la Investigación y la Tecnología}'' (Grant No. 5157). J. S.-V. is grateful to the University of Pittsburgh for its kind hospitality during his sabbatical year. This research was supported in part by the University of Pittsburgh Center for Research Computing through the resources provided. Specifically, this work used the HTC cluster, which is supported by NIH award number S10OD028483.

%
\bibliography{/home/jereson/PAPERS/Bib/Bibliografia.bib}
\end{document}